\documentclass[12pt]{book}
\usepackage[utf8]{inputenc}
\usepackage{graphicx}
\usepackage{amsmath}
\graphicspath{{./images/}}
\usepackage{setspace}
\usepackage[compact]{titlesec}
\usepackage[english]{babel}
\usepackage{float}
\usepackage{tabto}
\usepackage{csquotes}
\usepackage{multicol}
\usepackage[left=1in,right=1in,top=1in,bottom=1in]{geometry}
\usepackage[
backend=biber,
style=numeric,
sorting=ynt
]{biblatex}
\AtEndPreamble{\defbibheading{bibliography}{\section*{\normalfont\refname}}}
\graphicspath{{images/}}

\titleformat{\chapter}[display]
{\singlespacing\bfseries\Huge}
{\titlerule\filright\Large\chaptertitlename\ \Large\thechapter}
{0pt}
{\filright}
[\titlerule]
\titlespacing*{\chapter}{0pt}{0pt}{*2}

\begin{document}
    \thispagestyle{empty}
    \begin{center}
        \centering
        \textbf{Comprehensive Efficiency Analysis of Machine Learning Algorithms for Developing Hardware-Based Cybersecurity Countermeasures}\par
        \vspace{3cm}
        A THESIS\par
        Presented to the Department of Computer Engineering and Computer Science\par
        California State University, Long Beach\par
        \vspace{3cm}
        In Partial Fulfillment\par
        of the Requirements for\par
        the University Honors Program Certificate\par
        \vspace{3cm}
        By Darren Cobian\par
        \vspace{1cm}
        Fall 2021\par
        \vspace{1cm}
    \end{center}
    
    \newpage
    \thispagestyle{plain}
    \begin{center}
        \uppercase{Abstract}\par
        \textbf{Comprehensive Efficiency Analysis of Machine Learning Algorithms for Developing Hardware-Based Cybersecurity Countermeasures}\par
        
        By\par
        \vspace{-1ex}
        Darren Cobian\par
        \vspace{-1ex}
        Fall 2021\par
    \end{center}
    \vspace{-3ex}
    Modern computing systems have led cyber adversaries to create more sophisticated malware than was previously available in the early days of technology. Dated detection techniques such as Anti-Virus Software (AVS) based on signature-based methods could no longer keep up with the demand that computer systems required of them. The complexity of modern malware has led to the development of contemporary detection techniques that use the machine learning field and hardware to boost the detection rates of malicious software. These new techniques use Hardware Performance Counters (HPCs) that form a digital signature of sorts. After the models are fed training data, they can reference these HPCs to classify zero-day malware samples. A problem emerges when malware with no comparable HPC values comes into contact with these new techniques. We provide an analysis of several machine learning and deep learning models that run zero-day samples and evaluate the results from the conversion of C++ algorithms to a hardware description language (HDL) used to begin a hardware implementation. Our results present a lack of accuracy from the models when running zero-day malware data as our highest detector, decision tree, was only able to reach 91.2\% accuracy and had an F1-Score of 91.5\% in the form of a decision tree. Next, through the Receiver Operating Curve (ROC) and area-under-the-curve (AUC), we can also determine that the algorithms did not present significant robustness as the largest AUC was only 0.819. In addition, we viewed relatively high overhead for our ensemble learning algorithm while also only having an 86.3\% accuracy and 86\% F1-Score. Finally, as an additional task, we adapted the one rule algorithm to fit many rules to make malware classification understandable to everyday users by allowing them to view the regulations while maintaining relatively high accuracy.
    
    \newpage
    \thispagestyle{plain}
    \begin{center}
        \uppercase{\textbf{Acknowledgements}}
    \end{center}
    \tab I would like to thank my mentor, Dr. Hossein Sayadi, for keeping me focused on our task at hand and providing me with the necessary knowledge to complete this research. I would also like to thank my honors advisor, Liza Bledsoe, for keeping me motivated throughout this process. Lastly, I would like to thank my friends and family for supporting me through earning my Bachelor's degree.
    
    \thispagestyle{plain}
    \let\cleardoublepage\clearpage
    \renewcommand\contentsname{\centerline{\Large{Table of Contents}}}
    \tableofcontents
    
    \newpage
    \thispagestyle{plain}
    
    \renewcommand\listtablename{\centerline{\Large{List of Tables}}}
    \listoftables
    
    \newpage
    \thispagestyle{plain}
    
    \renewcommand\listfigurename{\centerline{\Large{List of Figures}}}
    \listoffigures
    
    \newpage
    \pagenumbering{arabic}
    \setcounter{page}{1}
    \thispagestyle{plain}
    
    \chapter{\large{Introduction}}
    
    \tab\\ \tab\tab  The security threats that emerging cybersecurity attacks bring on computer systems grows exponentially as the number of users on the internet steadily increases \cite{JANGJACCARD}. In addition, adversaries are using newer and more advanced attacks on computer systems as security experts continue to make strides in computer defense. Cybersecurity has always been a critical challenge to address in modern computer systems and information technology infrastructures. According to recent security reports, malicious software (a.k.a. malware) is rising at an alarming rate, and  the security of computing systems must keep up with the increase \cite{McAfee}. The attackers have leveraged emerging software and hardware vulnerabilities to compromise the security of systems and deploy malicious activities \cite{kocher2019spectre, lipp2018meltdown, mwscas_2020recent, wang2020comprehensive}. 
    
    Malware is the general term for malicious programs that damage or disable computer systems, networks, and mobile devices or leak sensitive information without the users' consent. Originally, attackers created malware to evolve or become metamorphic, polymorphic, and elusive as a way to circumvent anti-virus software and the signature-based methods that they used for detection \cite{Robust}. In an attempt to avoid modern detection software, adversaries formulate unique malware samples because they do not have existing data. This uncommon malware endangers user information that exists on the IoT along with its popularity and the many edge devices it employs; therefore, the demand for an advanced detection technique is increasing alarmingly \cite{Kwan, IoT}.
   
    These advancements have allowed for recent computer security defenses to delve into the world of machine learning to boost detection rates for various types of malware. Hardware-based Malware Detection (HMD) methods have recently emerged as an efficient approach for detecting malware at the processors' microarchitecture level with the aid of machine learning algorithms applied on Hardware Performance Counter (HPC) data \cite{Demme-ISCA13,Tang, sayadi_ensemble_2018,sayadi_2smart:_2019}. HMDs, help address the high complexity and computational overheads of conventional software-based malware detection techniques. HPCs are special-purpose registers designed in modern microprocessors to capture the hardware-driven events of running applications \cite{Rootkit-Singh,sayadi2021towards}. Hardware-supported malware detection methods have shown the suitability of standard machine learning (ML) algorithms applied on HPCs information to detect malicious application patterns. 

    Therefore, the shortcomings of defense in cybersecurity have led engineers to tread into undiscovered areas; one such area is hardware-based malware detection. These types of defense mechanisms could lower the overhead of algorithms in terms of their computational cost. However, to bypass this new hardware security, adversaries employed new methods of attack, and this time, the attacks would more specifically target hardware on a computer system \cite{Coin}. This cat and mouse game seems to be a never-ending battle. As the internet grows into spaces such as cities based on the IoT, new adversaries with new attack patterns will appear, and new defense methods will emerge to counter them \cite{micro}. At this point, defenses would once again become more advanced. However, along with these changes, the models' ambiguity would increase, meaning users do not know precisely how files are classified \cite{explain, Interpred}. Therefore, analyzing the efficiency and explainability of machine learning algorithms adopted for hardware-based cybersecurity countermeasures (e.g., malware detection) have become an emerging topic to explore. 
    
    
    Cybersecurity will always be a fascinating topic in the computing world due to the nature of the complexity of the attacks that adversaries take against users. The more critical security needs to get with more vital data going onto a cloud or an ordinary hard drive. It becomes a matter of computer ethics; computer scientists must build software with privacy and anonymity in mind to keep everyday users safe. Of course, this is the most crucial factor when creating cybersecurity software that protects against advancing malware threats. Anti-virus software, machine learning, and deep learning techniques help differentiate benign samples from malicious ones. However, recent studies found that machine learning based detection techniques outperform existing anti-virus software. 
    
    This thesis focuses on multiple contributions to machine algorithms applied for hardware-based cybersecurity against malicious software. We attempt to implement a hardware-based security countermeasure by using a wide range of machine learning algorithms 
    when detecting zero-day malware. 
    Our evaluations are in the form of a comprehensive efficiency analysis across different performance metrics such as detection accuracy, precision, recall, F1 score, Area Under the Curve (AUC), and expand up to the hardware resources used and latency from a single run. We collected these values from our proposed hardware implementation through Vitis HLS pipeline and were able to evaluate the implemented machine learning algorithms at a hardware level. 
    Additionally, we made a contribution to explainable machine learning by implementing our own One Rule (OneR) algorithm for hardware-based malware detection that changes the amount of rules made in an attempt to increase the detection accuracy in an explainable manner. This informative approach provides users with understanding of ML algorithms through interpretable machine learning. Lastly, we will evaluate our machine learning algorithms that have been converted into a hardware descriptive language (HDL) to decrease the computational overhead of the algorithms.
    
    
    This research thesis will be structured as follows. Chapter 2 will contain the background information focusing on the types of malware tested and related works to our topic. Chapter 3 will focus on the methodology, which includes an explanation of the data set, a description of the feature engineering done on the data set, a basis for each machine learning algorithm used, and an introduction to our attempts to accelerate these algorithms through hardware-based methods. Chapter 4 presents the metrics used for our model evaluations, the results produced, and the analysis of the hardware implementation. The final chapter will hold our closing remarks which include our future direction and conclusion.
    
    \newpage
    \thispagestyle{plain}
    
    \chapter{\large{Related Work and Background}}
    
    \section{\large{Hardware Performance Counters}}
        
    \tab Hardware Performance Counters (HPCs) are a list of values retrieved from an operating system's internal registers and are used to replace signature-based malware detection. HPCs monitor micro-architectural events (e.g., node-loads, dTLBstores, CPU/branch-instructions, cyclesct) while an application is executing \cite{HPC1,sayadi_ensemble_2018,wang2020comprehensive}. The values are collected to form a pattern that can uniquely identify an event. 
    Every processor has its own set of HPCs initially used for optimization and performance tuning; on a Linux system, its built-in \textit{perf} tool can retrieve these values for us \cite{sayadi_2smart:_2019}. The introduction of metamorphic malware caused signature-based methods to become outdated. Therefore, achieving higher accuracy on metamorphic or zero-day malware required modern detection techniques to classify the more advanced HPCs \cite{Coin, Robust}. The initial purpose of performance counter registers is to analyze architectural level performance and power of applications \cite{sayadi2017-igsc,makrani2018energy}. While HPCs are finding their ways in various processor platforms from high-performance to low power embedded processors, they are limited in the number of micro-architectural events that can be captured simultaneously. 

    The HPCs that express a malicious or benign sample can be obtained both before and after various files containing malware had run through a Linux system. We then use these values to train numerous machine learning and deep learning (e.g., logistic regression, deep neural networks) algorithms. The model that returns the highest accuracy from classification can be used to test new HPC values or zero-day samples for real-time detection. Since most HPC retrievals return with more than a few features columns, feature selection/engineering needs to be done on the data before a model can properly use the HPC data set for training \cite{PMC}.
    
    \section{\large{Related Work}}
    
    \tab The reliability of a system is determined by its ability to secure the anonymity of a user by protecting their critical information from malicious attacks. Numerous attack styles target weak points on a system to gain a user's data. The work in \cite{cache} introduces cache side-channel attacks, an attack that tracks a user's activity and, over time, gathers enough information to learn something about the person they are following. The attack does this by saving tidbits while a specific program is active on the system.  Attacks work in two stages: the first requires a cache block to be set to a particular state before the victim executes a program. Then, after a program runs, the attacker gathers memory traces over multiple iterations, and once done enough times, the adversary can build  confidential information about a user \cite{cache}. The work in \cite{wang2020mitigating} introduced a defense mechanism which does not require a modification of a processor to mitigate attacks in the form of a randomization technique. Wang et al. implemented this method by using a processors frequency and varying its setting to create an obstruction that would in turn stop the leakage of critical data. This paper offered a lightweight system and architecture level randomization technique to effectively mitigate the impact of side-channel attacks on last-level caches with no hardware redesign overhead. They adapted the processor frequency and prefetchers operation and added proper level of noise to the attackers’ cache observations to prevent the critical information from being leaked. Zhang et al., on the other hand, proposed deep neural networks such as multilayer perceptrons (MLP) to test a system's defenses. To train the MLP, the authors use the observations gained from each attack. \cite{cache} tested their model on newly gathered memory traces and the results contain a victim's execution information. The training set is modeled as a state machine to model key characteristics of different cache attacks \cite{cache}. They then analyze the accuracy produced by the MLP. By evaluating the accuracy based on the path from memory to execution traces, they could determine exactly how much of the retrieved data would be correct on any given attack. \par 
    
    Machine learning software has grown to complex levels for the defense of modern computer systems. One such subfield is deep learning which contains neural networks metaphorically compared to the human brain. It includes algorithms such as deep belief networks, recurrent neural networks, convolutional neural networks, and at its simplest form, multilayer perceptrons \cite{ML_DL}. Unfortunately, modern attacks opt to attack the test set that runs through the algorithm after training because they use the data to make it look like a simple malware sample instead of a benign one. However, by taking advantage of deep learning, computer systems can protect against obfuscated malware \cite{Defense}. Kuruvila et al. counter this approach by proposing a moving target defense (MTD) that constantly changes the attack surface. They built the MTD on multiple algorithms and trained each algorithm on a different set of HPCs which means that an adversary could never pinpoint the HPC values and change them to allow for their malware to bypass a system \cite{Defense}. This cat and mouse game seems to be a never-ending battle. As the internet continues to grow into spaces such as cities based on the IoT, new adversaries with new attack patterns will appear, and new defense methods will appear to counter them. \par
    
    After analyzing complex and simple machine learning models, algorithm descriptions used the terms "Eager Learner" and "Lazy Learner" (instance based) multiple times to describe specific algorithms \cite{instance}. \cite{Lazy_Eager} covers these two topics by defining the following algorithms: linear regression, multilayer perceptron, random forest, K-Nearest Neighbor (KNN), weighted learning, support vector machine, J48, decision tree, and KStar. They state that the eager method requires less space than the lazy method. They used the algorithms to classify rent predictions and explained that the volume of the data did not matter in the case of eager learning. On the other hand, lazy learners aimed to find local optimal solutions for each test input in a data set. They state that the data set matters in both eager and lazy methods, and therefore they focus on feature selection before designing the algorithms. Once these steps are complete, the algorithms could be classified. Eager learners learn from the training data before they read test data. A few examples of eager learners include multilayer perceptrons, random forests, linear regression, J48, support vector machine (SVM), and SMO. Locally weighted learning, KStar, Lazy-Decision Tree ML-KNN, and KNN all use the entire data set for training and testing and are considered lazy learners.
    
    In this section, we discuss related work for machine learning solutions for hardware-based malware detection. Demme et al. \cite{Demme-ISCA13} was the first study to examine the effectiveness of hardware performance counter information for accurate malware detection. The authors proposed using hardware performance counter data to detect malicious behavior patterns using machine learning techniques primarily on mobile operating systems such as Android. The paper ultimately successfully demonstrated the effectiveness of offline machine learning algorithms in identifying malicious software. In addition, it illustrated the suitability of employing HPC information in detecting malware at the Linux OS level, such as Linux rootkits and cache side-channel attacks on Intel and ARM processors. Finally, it exhibited high detection performance results for Android malware by applying complex ML algorithms, namely Artificial Neural Network (ANN) and K-Nearest Neighbor (KNN). 

    In a different study, Tang et al. \cite{Tang}  further discussed the feasibility of unsupervised learning that employs low-level HPCs features for detecting return-oriented programming (ROP) and buffer overflow attacks by finding anomalies in hardware performance counter information. This study used the Fisher Score metric to identify the top 7 low-level features for malware detection for feature selection. These reduced features are then used to train unsupervised machine learning methods to detect program behavior deviations from a potential malicious attack. The work further provides a comparison of performance using different sampling frequencies of the HPCs. 

    The work in \cite{sayadi_ensemble_2018}  has examined the use of ensemble learning and standard machine learning models to realize an effective run-time hardware-based malware detection with a limited number of performance counters available in modern microprocessors.  The two techniques they used were boosting and bagging. Boosting is a technique that weighs each training data set and adjusts the weights based on the model's overall accuracy. On the other hand, Bagging is another ensemble learning technique that takes a statistical value from multiple random samples and uses it to train the ML models. They compared the robustness and the accuracy of regular classifiers with boosted classifiers and found that boosting techniques improved the classification by as much as 17\% with a much lower amount of HPCs.\par  
    
    In addition, recent work in \cite{sayadi_2smart:_2019} proposed a two-stage machine learning-based approach for run-time malware detection. The first level classifies applications using a multiclass classification technique into either benign or one of the malware classes (Virus, Rootkit, Backdoor, and Trojan). In the second level, to have a high detection performance, the authors deploy an ML model that works best for each class of malware and further apply compelling ensemble learning to enhance the performance of hardware-based malware detection. The work in \cite{sayadi2018customized} also proposed an effective machine learning-based hardware-assisted malware detection framework for embedded devices that only utilizes a limited number (only 4) of low-level features in a microprocessor, i.e., HPC events, to facilitate the run-time malware detection. 

    Singh et al. \cite{Rootkit-Singh} is a recent work on HMD that deploys machine learning algorithms applied on synthetic traces of HPC features to detect kernel-level rootkit attacks. For feature reduction, they process the application traces using the Gain Ratio feature selection technique from the WEKA machine learning toolkit to determine which features are the most prominent for each rootkit. As a result, the authors achieve high prediction accuracy in detecting five self-developed synthetic rootkits models. Nevertheless, while necessary, this work only focused on detecting kernel rootkit attacks using a limited set of synthesis data sets. 

    The authors in \cite{Sayadi-glsvlsi20, sayadi2021towards} addressed the challenge of detecting embedded malware using hardware features. Embedded malware refers to harmful stealthy cyber-attacks in which the malicious code is hidden within benign applications and remains undetected by traditional malware detection approaches. However, in HMD methods, when the HPC data is directly fed into an ML classifier, embedding malicious code inside the benign applications leads to contamination of HPC information, as the collected HPC features combine benign and malware microarchitectural events. To address this challenge, the authors present StealthMiner, a specialized time series machine learning approach based on Fully Convolutional Network (FCN) to detect embedded malware at run-time using branch instructions feature, the most prominent HPC feature. 

    Moreover, recent research in \cite{adaptive-HMD} conducted analysis across a wide range of malicious software applications and different branches of machine learning algorithms and indicated that the type of adopted ML algorithm to detect malicious applications at the hardware level highly correlates with the type of the examined malware and the ultimate performance evaluation metric to choose the most efficient ML model for distinguishing the target malware from the benign program. Therefore, they presented \textit{Adaptive-HMD}, a cost-efficient ML-driven framework for online malware detection using low-level HPC events. This method is based on a lightweight tree-based decision-making algorithm that accurately selects the most efficient ML model for the inference in online malware detection according to the users' preference and optimal performance vs. cost criteria. The experimental results demonstrate that Adaptive-HMD achieves up to 94\% detection rate (F-measure) while improving the cost-efficiency of ML-based malware detection by more than 5X compared to existing ensemble-based malware detection methods.

    Standard detection techniques struggle in identifying zero-day malware samples while also managing a high accuracy \cite{ISQED}. The work compared six ML-based HMD classifiers and found that when they tested most algorithms with zero-day data, accuracy was reduced by almost 30\%. To handle this challenge when testing zero-day samples,\cite{ISQED} proposed an ensemble learning-based technique to enhance the performance of the standard malware detectors. The \textit{perf} tool from a Linux kernel system gathered the HPC values from the operating systems' registers, and only the top four features were run through their classifiers \cite{ISQED}. Initially, the file contained 16 features that would reduce the performance of the algorithms if run in this state \cite{ISQED}. They managed to choose the top 4 features using recursive feature elimination (RME) provided by SciKit Learn. The authors also explained how they are testing their proposal. They used Random Forest from the standard ML-based detectors because it produced the most significant accuracy. They tried their random forest implementation using 10-fold cross-validation and zero-day data and then fed it to AdaBoost, a boosting classifier. They found that the ROC AUC score increased by almost 5\% and broke into the 10th percentile. In addition, they noted that their proposal had a 95\% TPR while only having a 2\% FPR, and their model provided negligible latency.\par
    
    Modern ML-based techniques opt to use HPC registers for malware detection. This is because signature-based approaches are often exploited and generally have had lower success rates \cite{PMC}. \cite{HPC1} compares to different HPC approaches, namely ready-made and tailor-made HPCs. In particular, Kuruvila et al. state that the main difference between the two is based on granularity. Ready-made HPCs exhibit low granularity, which inhibits a complete classification instead of tailor-made HPCs with high granularity. Additionally, tailor-made HPCs count assembly-level instructions at run-time, which allowed for a higher level of defense against side-channel exploits and resulted in almost a 10\% increase in accuracy when run through a Random Forest classifier. Therefore, it \cite{HPC1} determined that Tailor-made HPCs were superior and capable of securing sensitive information while also detecting anomalies in real-time.\par
    
    So far, our understanding of machine learning has come through examples specifically modeled toward malware detection. However, it can also be used in other fields such as cyber-security, as discussed by \cite{ML_DL}. The models could be used to detect misuse, anomalies, and intrusion while both on a wired and wireless connection, as explained by \cite{DEPREN}. They state that the data set used to train the models for this proposal will be essential for detecting network security issues. They used a support vector machine (SVM), K-NearestNeighbor (KNN), Decision Tree, Deep Belief Network, Recurrent Neural Networks, and Convolutional Neural Networks. The authors described issues that can lead these models to fail and how they can be better. For example, they stated that detection speeds can be increased for machine learning and deep learning algorithms and could use hardware in parallel to raise rates. Additionally, they note that data sets can quickly become outdated, so programmers should always gather new data sets for model building. \par
    
    Signature-based malware detection methods have become outdated; the way it works is that Anti-Virus Software (AVS) reads values from a software application and tries to match this 'signature' to existing malware signatures. However, as stated in \cite{Detection}, detection software can classify malware into worms, Trojans, bots, viruses, etc., and these malware types can easily overcome AVS signature-based detection through obfuscation methods. Due to this, Yuxin et al. proposed a new detection method based on Deep Learning, a sub-field of machine learning. Their Deep Belief Network (DBN) accepts unlabeled data and topology of multiple hidden layers to achieve high accuracy. The data passed into their model are opcode n-grams extracted from Portable Executable (PE) files by a PE Parser. Before the DBN accepts the n-grams, the authors used a feature extractor that created a data set containing only the necessary n-grams, leading to higher accuracy. Finally, they evaluated their classification results using their accuracy and F1 scores. They compared them to other machine learning algorithms such as a support vector machine (SVM), K-Nearest-Neighbor (KNN), and Decision Trees. They received successful results and, after comparison to these algorithms, proved to have better performance. \par
    
    Because of its robust nature, deep learning has been seen as the answer to protect against malware that becomes obfuscated through metamorphism. \cite{Robust} proposes a solution named ScaleMalNet which uses image processing for malware classification. The general cycle for this proposal has the first step listed as classification of a file either being malware or benign sample. The second step gives a more detailed approach to telling the user exactly what type of malware the infected file contains. Finally, the model returns an evaluation chart that allows us the understand its accuracy on the given data. A base case for classification in their model comes from malware that has similar if not the same image as a previously classified file if it does return that type of malware and end the program. It is important to note that the pictures come from converting binary malware into greyscale images and are then fed into the model. ScaleMalNet uses a deep learning image processing approach to prevent metamorphic malware from getting through its defenses. The authors use a DL model with 4608 hidden layers and ten output nodes with an Adam Optimizer and a softmax activation function with categorical-cross entropy for their loss function. This approach on the model led to a classification accuracy of 96.3\%, which was a resounding success. \par 
    
    Many models could have a significant overhead when it comes to detecting malware. The research in \cite{Security} attempts to lower the cost of detecting malware through their proposal, SPIREL. The solution they propose uses an organization's priorities and requirements and, based on this, automatically chooses the most computationally efficient detector. SPIREL decides where a deep neural network has enough data to make a classification. The authors use a reinforcement learning model which self trains based on rewards and loss from correct and incorrect classifications. If the prediction is correct, the model receives compensation, but if the model is wrong, the model takes a loss. The model's efficiency is determined; a fast run time means that the model is efficient with a large amount of data. As in the previous paper, PE and APK file n-grams train the model and test the accuracy. The model had a success rate of 73\% on PE files and 78.6\% on APK files. Reinforcement learning is used here for scheduling and optimization, and as mentioned, it does this by immediately choosing a classifier as soon as the model feels it is ready. Finally, SPIREL can easily manage its cost by adjusting its policy whenever it encounters a false positive or a false negative. \par
    
    Conversely, as security continues to grow in favor of detecting malicious software, adversaries are also attempting to find ways around our security techniques. For example, it \cite{Coin} introduces Hardware Trojans (HT) and counterfeit Integrated Circuit (IC), which can get by the DBN as mentioned earlier. They explain that HTs are used for chip malfunctions and the compromising of sensitive information, while counterfeit ICs compromise and jeopardize the reliability of a system. The goal of the Trojan is to leak data, and it's able to get onto an IC by tampering done by a manufacturer, and the authors explain that they are typically placed off the main circuit path to avoid detection. To detect both HTs and counterfeit ICs, on-chip data analysis, their SVM takes advantage of gate-level netlist analysis and run-time traffic information. Deep Neural Networks (DNN) are not always the best classification model, and as shown here, they went to a less accurate model for detection. Instead, they are now feeding the SVM a new type of data for classification results. They state that counter-measures for hardware attacks include limiting error by model enhancements, building error-resilient hardware to mitigate hardware fault attacks, and using root-of-trust to safeguard DNN IP cores and sensitive information. \par
    
    Understanding how a machine learning algorithm works when used as a classifier through hardware proves to be complicated. The results are straightforward but offer no explanation on how a sample is classified because users are unaware of what is happening under the hood; in their eyes, it just seems to be working. Authors in \cite{explain} present an explainable approach to using machine learning for hardware-based malware detection. This approach uses HPC counters to formulate a data set, but other methods introduce confusion because the result remains binary without explaining their data set or the steps taken to get there. Their goal is to perform outcome interpretation. Pan et al. first step in doing this is by giving the essential features more significant coefficients than the rest in their classification formula. The most critical features, in this case, would be those that lead to positive detection of malware. They also check the clock cycle distributions of these top features to determine when exactly malware was present by comparing them to when Linux ran specific files or software. It is worth mentioning that their data set was made up of trace values that, when looked at by individual cells, can produce enough information to detect malware and clock cycles and large coefficients. The mixed results led to an explainable model which possessed an accuracy greater than existing classifiers such as PREEMPT, which uses decision trees and random forests for classification; their model constantly had an accuracy greater than 90\% and had a low false-positive rate. \par
    
    Modern malware use obfuscation techniques to overcome AVS, forcing hardware-based malware detectors to come to the forefront of classification techniques. Just as defending against malware becomes an evolutionary process, so do adversaries' attacks against the malware sensors. The work in \cite{micro} presents a new attack style that uses the HPCs to detect malware through hardware to conduct their attack. They use machine learning to create this attack type, and more specifically, Dinakarrao et el. use logistic regression, a decision tree, Naive Bayes, and neural networks to simulate it. The classifier that produced the highest accuracy would be the algorithm assumed to be used by an attacker. To get this point, they first needed to understand where the attacker might focus their attack, which meant the first step was finding their weakness or flaw. They state that an adversarial sample generator is wrapped around a standard application to produce HPC patterns similar to an adversarial trace. By doing this, they created a malicious file that had no malicious behaviors, which meant that it would be hard to detect. After feeding this to their highest accuracy model at 82.76\%, their accuracy was reduced by 18.04\%.
    
    \raggedbottom
    \section{\large{Different Types of Malware}}
        
    \tab The scale that malicious software (malware) works at can be a frightening endeavor to take on for computer scientists in the cyber-security field. They can come in various shapes and sizes and can come either directly from a piece of downloaded software or be rooted in your hardware, where it will activate upon the occurrence of a particular computational event \cite{Security}\cite{micro}\cite{Coin}. Adversaries use zero-day malware to gain sensitive information or to bring harm to a device. They accomplish this by gathering data over time without your knowledge of the malicious software even being present on your device or by creating a vulnerability in your computer's security system \cite{Security}. This thesis treats malware as the big picture; however, our data set contains nine separate entities, each with unique HPC equivalent samples whose target class is labeled "malware."\\
    
    \begin{table}[H]
    \caption{\large{Different Types of Malware}}
    \begin{center}\scriptsize
        \begin{tabular}{| l | p{13cm} |}
        \hline
            \textbf{Malware} & \textbf{Description} \\ \hline
            Backdoor &  Backdoors cause a vulnerability on a device by leaving a "door" open for adversaries to gain root access or control of a system. Root control gives an adversary access to everything on a system. Examples: Design and Hardware/Software\\ \hline
             Worm & The internet first encountered worms in 1988. They can propagate through different shared media sources such as USB drives, and so long as they are attached to a piece of software, they will spread. Example: Morris Worm \cite{Robust}\\ \hline
             Virus & Unlike a worm, a virus is not required to be attached to a piece of software to spread. Instead, a virus propagates once the dependent software it is attached to executes. It immediately infects the next piece of software it finds. It is worth noting that a worm is a virus, but a virus is not necessarily a worm. \\ \hline
            Rootkit & Rootkits are difficult malware to detect, as are the processes they attempt to hide from a computer system. \\ \hline
            Botnet & Botnets are a type of malware attack that takes place over the internet. If successful, an adversary can take control of a system while in a remote location. The Internet-of-Things (IoT) sensors are a common area of attack \cite{Botnets} \\ \hline
            Ransomware & Ransomware is a type of malware that attaches to your computer, finds sensitive information, and keeps it hostage. At the same time, the adversary behind the attack demands compensation for the safe return of your private information.\\ \hline
            Spyware & Spyware is a type of malicious program that attaches to your computer and silently tracks your every move. Including web searches and passwords used \cite{spyware}. \\ \hline
            Adware & If you've worked with computers more than once, chances are you have most likely encountered adware in one way or another. It is a type of malicious software that downloads and displays advertisements while you are on the internet. \\ \hline
            Trojan & Similar to botnets and backdoors, once this malware is activated, it gives an adversary full access to your computer system. Usually well disguised and difficult to encounter. Trojans require a trigger before they begin to work maliciously. \\ \hline
        \end{tabular}
    \end{center}
    \end{table}
    
    \newpage
    \thispagestyle{plain}
    
    \chapter{\large{Methodology}}

    \section{\large{Feature Engineering}}
    
    \tab The dataset used in our implementations comprises values gathered from HPCs from when various benign and 
    malware samples were running on a Linux system \cite{ISQED}. The Linux 4.4 Kernel was running on an Intel Xeon X5550 machine on the Ubuntu 14.04. Since the dataset is contrived from hardware registers, we use the \textit{perf} tool from a Linux operating system to return the necessary HPC values. 
    To create the training and test sets, a 80\% split was done that left exactly five malware samples in the training set and the remaining 20\% contained four zero-day malware samples. Specifically, the 5 types of malware that are included in the training set are different from the 4 types contained by the test set. This is due to a detection technique not always having a certain type of malware's existing data. To that end, this means that we must evaluate the metrics from an algorithm based on how accurate or robust a program is against zero-day malware samples. Before we split the data, it is necessary that we use feature selection to decrease the size of the data set. Otherwise, the sheer amount of data that would go into the model would have a massive overhead and slow down our classifications.
    
    As mentioned, running these large datasets combined would take far too long and lead to the inefficiency of the models; therefore, we must introduce feature engineering and ease the load that the models take on. We take advantage of SciKit-Learn, a library that allows us to use a variety of machine learning algorithms by calling its methods. A few of those methods include the ability to significantly reduce a data set down to only its most essential features. Specifically, we use "SelectKBest," which takes a function to score the feature vector and a number k whose value totals the number of features we want to keep; in our case, we observed four features.  The higher the score, the more dependent a feature is on another, and therefore the data set will keep those with the four highest scores. The resulting features were node-loads, dTLBstores, branch-instructions, and cyclesct. \par
    
    Along with "SelectKBest," Mutual Information is used to help select the top features. It returns a score based upon the ability of one feature to explain another in the same data set; the mathematical approach is presented in equation 3.1. The function itself also comes from sklearn's "feature\_selection" library and is titled "mutual\_info\_classif." It works in tandem with "SelectKBest" to return a new NumPy array that contains all the necessary columns and targets. More specifically, the score that it produces is a probability-based on dependencies from column to column that allows us to speed up the learning process \cite{KIRA}. The formula can be written as a probability mass function as follows:
    
    \begin{equation}
        I(X;Y) = \sum_{x\epsilon X, y\epsilon Y} p_{(X,Y)}(x,y)log\frac{p_{(X,Y)}(x,y)}{p_{X}(x)p_{Y}{(y)}}
    \end{equation}
    
    \begin{figure}[H]
        \begin{center}
            \hbox{\hspace{9ex} \includegraphics[scale=.8]{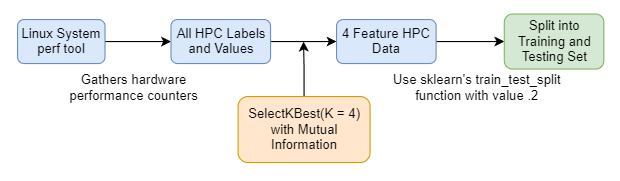}}
            \caption{\large{Feature Selection Chart}}
        \end{center}
    \end{figure}
    
    \section{\large{Classification on Zero-Day Malware Detection}}
    
    \tab Holes left by the rapid development of modern-day computing have increased the challenge of defending against malicious software \cite{sayadi2021towards}. With adversaries finding these new exploits, new malware finds its place in cyberspace. These new variants have forced new detection techniques to come forward to make the internet a safer place, but overhead becomes an issue when malware begins to attack edge technology. These edge devices typically belong to the IoT. Their complexity can make it difficult to detect malware such as trojans that wait for a specific event before they make their move \cite{rajatTrojan}. This problem tied with zero-day malware being challenging to detect means that finding an effective malware detection solution will have a colossal impact in shaping the future of the internet\cite{ISQED}. 
    
    A total of seven different classifiers were used for our experiments; they included KNN, OneR, MLP, bagged tree, decision tree, logistic regression, and SVM. OneR and MLP were visualized through the WEKA data mining tool and reverse engineered through its output. Bagged tree, decision tree, logistic regression, and SVM were written in MATLAB and converted into C through the MATLAB Coder app. KNN was written entirely from scratch and utilized euclidean distance to achieve results. The explanations of these algorithms will be in the following order: (3.3.1) KNN, (3.3.2) MLP, (3.3.3) Bagged Tree, (3.3.4) Decision Tree, (3.3.5) Logistic Regression, and finally (3.3.6) SVM
    
    \subsection{\large{K-NearestNeighbor (KNN)}}
    
    \tab KNN is a lazy machine learning classifier that is based on a distance function. It measures the difference or similarity between two points \cite{ML_DL}. Two different metrics can approximate the measurement, but the most common are the Manhattan and Euclidean distances. The formulas for each are as follows. \newline
    
    \textbf{Manhattan Distance:} 
    \begin{equation}
        d = \sum_{i=1}^{n} \mid x_i - y_i \mid 
    \end{equation}
    \newline
    \tab \tab \textbf{Euclidean Distance:}
    \begin{equation}
        d(x,y) = \sqrt{\sum_{i=1}^{n} (x_i - y_i)^2}
    \end{equation}

    In this experiment's model, the Euclidean Distance is used, and points are created by comparing the test set values with the training set values. The summation, in this case, begins at 1 for i and ends at a value of n equals 29,941, which is the size of the training set. X and y denote a position at the ith row of the training set. New points can get classified by comparing them to the nodes created by the training set.

    KNN, in this case, uses binary classification, and a K value is fed to the algorithm usually as an odd number, so the nearest points located to a test feature vector will always have a majority target. The test values could then be appropriately classified into either benign or malicious samples \cite{ML_DL}. A simple visual representation of the KNN algorithm can be observed in Figure 3.2. \par
    
    \begin{figure}
        \begin{center}
            \hbox{\hspace{24ex} \includegraphics[scale=.8]{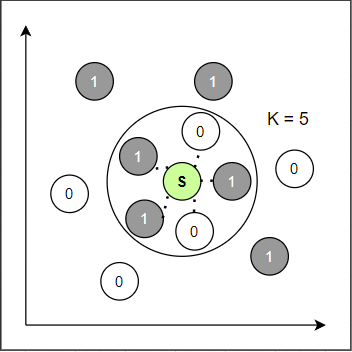}}
            \caption{\large{KNN Graphical Representation}}
        \end{center}
    \end{figure}
    
    We achieved an accuracy of 90.6\%, and in this model, the zero-day data yielded a precision of 90.4\%, a recall of 90.6\%, and a calculated F1-score of 90.5\%. Although the algorithm's accuracy was relatively high, the F1-score represents the data running quite well with our zero-day data.
    
    \subsection{\large{Multilayer Perceptron (MLP)}}
    
    \begin{figure}[!ht]
        \begin{center}
            \hbox{\hspace{22ex} \includegraphics[width=.5\textwidth]{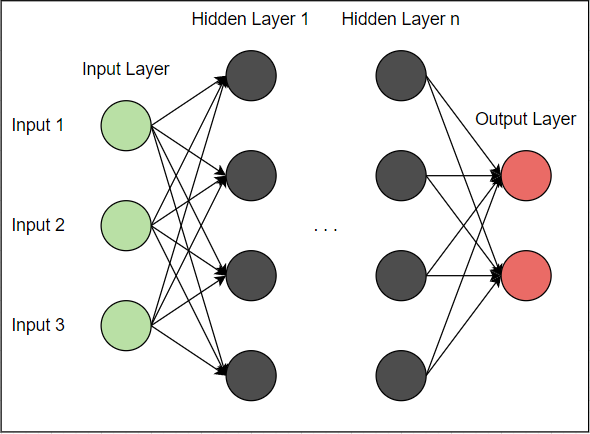}}
            \caption{\large{MLP General Model}}
        \end{center}
    \end{figure}
    
    An MLP is a deep neural network (DNN) that uses back-propagation and a feed-forward approach through a directed graph that could not and should not create a cycle \cite{Robust}. An MLP will always have three or more layers that consist of one input layer, one output layer, and one or more hidden layers \cite{Robust} and a general MLP can be viewed in figure 3.3. In our case, we only have three layers: one input layer, one hidden layer, and one output layer. Additionally, we have a bias node in our hidden layer that contains a value of 1 for our equation.\par
    
    \begin{figure}[!ht]
        \begin{center}
            \hbox{\hspace{8.5ex} \includegraphics[width=.8\textwidth]{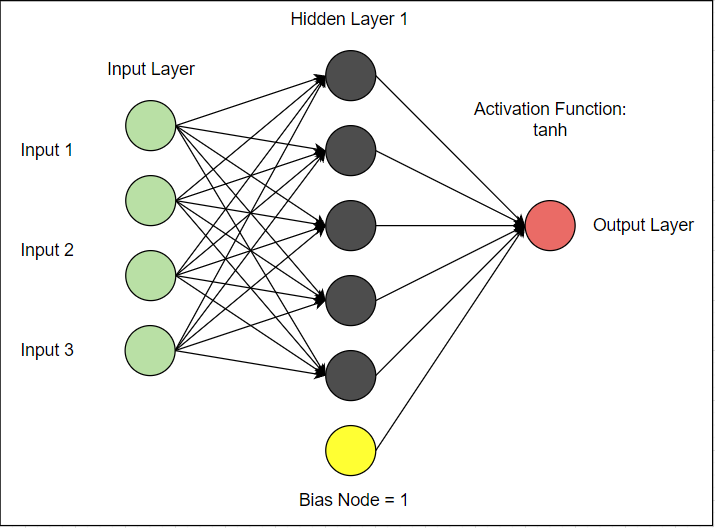}}
            \caption{\large{Our implemented MLP}}
        \end{center}
    \end{figure}
    
    Several steps must occur before the MLP can make a proper prediction. The first of which is reading in the values that will train the dataset, of which the final 6900 samples are from our test set, and by the time the MLP reaches these values, our accuracy should be much higher due to training. The first line in our data text file contains the topology of the sample, which in our case follows a 4 5 1 order. This topology order means that there are four nodes in our input layer, a total of 5 nodes in our single hidden layer, and our output layer contains only one output node; each of these nodes is interconnected and visual representation can be seen in figure 3.4. \par
    
    Before our MLP model can classify a new sample, a calculation must occur through the feed-forward network: \par
    \begin{equation}
        a_1 = w^Tx + b
    \end{equation}
    
    If we had instead created a DNN that contained multiple hidden layers, then in the above formula, $a_1$ would replace our x value if it would need to be passed forward to the next hidden layer or output layer. However, before our values move forward to the next hidden or output layer, it first needs to go through the activation function. Therefore, we opted to use tanh for our activation function, which is:
    \begin{equation}
        \sigma(z) = \frac{e^z - e^{-z}}{e^z + e^{-z}}
    \end{equation}
    
    The tanh function allows for an output between -1 and 1 where -1 will be equivalent to 0 or a benign sample, and one will be equal to 1 or a malware sample. At each layer, the MLP runs values through an activation function which allows the model to update its weights to ensure a more accurate prediction after the output layer. It takes advantage of the Root-Mean Squared Error (RMSE) and function gradients to help our model determine accurate weights: \par
    
    \begin{equation}
        RMSE = \sqrt{\frac{\sum_{i=1}^{N}(Predicted_i - Actual_i)^2}{N}}
    \end{equation}
    
    Once the weights at every level and node have been updated, the model runs through its final sample to get our final measurements. Accuracy was at 87.9\% by the end of our run while precision was at 77.3\%, the recall was at 87.9\%, and through these values, we got a calculated F1-score of 82.3\%. \par
    
    \subsection{\large{Decision Tree}}
    
    \tab Decision Trees are supervised learning algorithms usually used to create an algorithm based on explainable machine learning. In this case, we are using it as a predecessor to explain how a bootstrap aggregated algorithm or Bagged Trees work. \cite{Defense} defines a decision tree as an algorithm with a tree-like structure built upon mutually exclusive if-then rule sets. These if-then statements come from the malware and benign training dataset fed to the model, and typically, the most important rules are kept near the root of the tree. A general diagram of a decision tree is presented in Figure 3.5. \par
    
    \begin{figure}[!ht]
        \begin{center}
            \hbox{\hspace{8.5ex} \includegraphics[width=.8\textwidth]{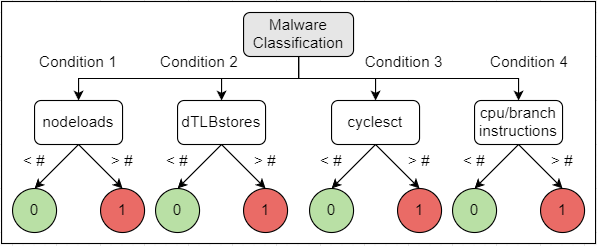}}
            \caption{\large{Our implemented decision tree}}
        \end{center}
    \end{figure}
    
    The decision tree used is based on the C4.5 algorithm and produced a model with an accuracy of 91.2\%. Next, we calculated the precision and recall of the model by gathering the false positives and negatives and the true positives and negatives. After these values were collected, we produced precision and recall of 91.7\% and 91.2\%, respectively. Finally, through these values, we calculated an F1-score of 91.5\%. \par
    
    \subsection{\large{Bagged Trees}}
    
    \tab Bagged Trees is an approach in machine learning that attempts to increase the accuracy of multiple types of models. In this case, we are trying to improve the accuracy of the previously covered Decision Tree model. Bagging refers to aggregating classifiers over bootstrap datasets \cite{rao1997visualizing}, that is, a dataset $\gamma$ is split into multiple smaller datasets $\beta_i$ that each have n values. It is possible that these n sample values be replaced in one of the other $\beta_i$ sets. Once this step is complete, the ensemble classifier creates multiple fine decision trees. The test set is then fed to the model and one sample is ran through each of the decision trees that it creates \cite{rao1997visualizing}. The model tallies the classification results from each decision tree, and we can interpret an example of what the results might look like from Figure 3.6. \par
    
    \begin{figure}[!ht]
        \begin{center}
            \hbox{\hspace{8.2ex} \includegraphics[width=.8\textwidth]{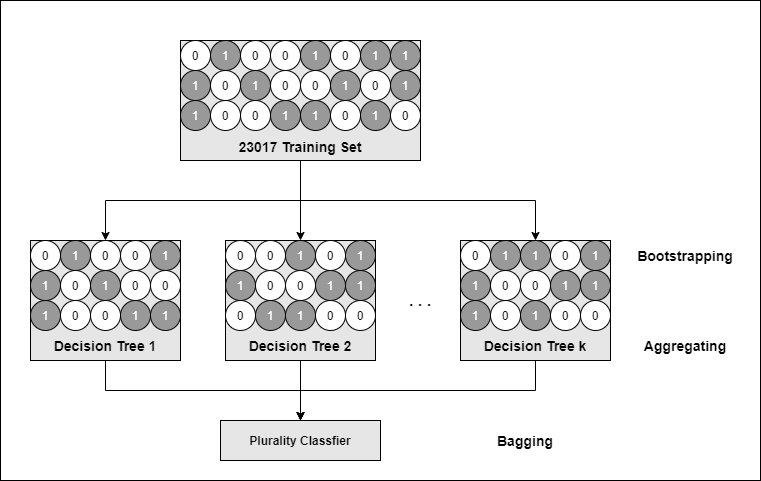}}
            \caption{\large{Bagged Trees Model}}
        \end{center}
    \end{figure}
    
    Once the ensemble algorithm reaches this stage, it can use a plurality rule to make a final decision on classifying a given sample. Bagged Trees have the potential to be more accurate than a simple decision tree because if a selection yields an unstable result, it can completely delete or erase the decision tree that caused it and thereby increase the accuracy of the following sample and the model as a whole \cite{rao1997visualizing}. \par
    
    \begin{table}[H]
        \caption{\large{Bagged Trees Aggregation Interpretation}}
        \begin{center}
            \hbox{\hspace{8.2ex} \includegraphics[width=.8\textwidth]{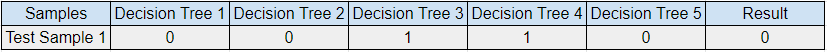}}
        \end{center}
    \end{table}
    
    This model leads us to our fifth-highest accuracy and our fourth-largest F1 score. These results yielded an accuracy of 86.3\%, a precision of 85.8\%, a recall of 86.3\%, and finally, a calculated F1-score of 86\%. The bagged trees algorithm did produce a higher accuracy than a single fine decision tree, as was its original goal, and therefore, the results of this model were a resounding success. \par
    
    \subsection{\large{Support Vector Machine (SVM)}}
    
    \tab A Support Vector Machine (SVM) is a type of supervised learning algorithm that classifies training samples with the help of a hyperplane defined as being greater than or less than a value of plus-or-minus 1 \cite{Coin}. More specifically, a formula based upon the weights, feature vector, and bias will be compared against the hyperplane; the hyperplane will separate the positive and negative sides of the model. The equations are as follows:\par
    \begin{equation}
        w^Tx + b = 0
    \end{equation}
    \begin{equation}
        \begin{cases}
            w^Tx + b \geq 1 \\
            w^Tx + b \leq -1
        \end{cases}
    \end{equation}
    We can then compare the graph of the model to the graph of a multilayer perceptron which changes as it updates its weights through back-propagation. The difference is that in an MLP, only one line is created, while in an SVM, a hyperplane is created where one line goes through positive one and the second line goes through negative one; it is worth noting that both lines will have an equal slope and this can be seen in figure 3.7. \par
    
    \begin{figure}[!ht]
        \begin{center}
            \hbox{\hspace{30ex} \includegraphics[width=.32\textwidth]{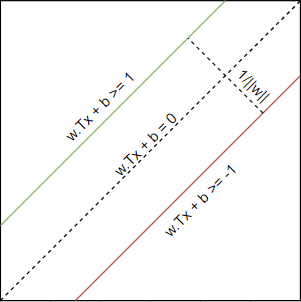}}
            \caption{\large{Simple SVM}}
        \end{center}
    \end{figure}
    
    The thickness or width of the hyperplane is determined by the magnitude of the weights of the model \cite{Coin}, as shown in equation 8, and if the weights are updated, so is the width of the hyperplane. We can also see that the midpoint of the hyperplane, as denoted by setting the equation equal to zero as in  equation 3.7, will always be directly in the center of the calculated hyperplane or graph. \par
    \begin{equation}
        Width = \frac{1}{\mid\mid w\mid\mid}
    \end{equation}
    \tab The SVM model produced an accuracy of 86.2\%, which ended up being our second least accurate classifier. Additionally, we calculated precision and recall of 85.6\% and 86.2\%, respectively. Finally, we calculated an F1-score of 85.9\% for our SVM model using the given precision and recall. \par
    
    \subsection{\large{Logistic Regression}}
    
    \tab Logistic Regression is one of the first two algorithms taught when learning about machine learning because of its low complexity. It is a supervised learning algorithm that uses feature vectors or input values and combines them linearly to make a binary prediction \cite{Coin}. Although logistic regression also uses an activation function, it is natural to choose sigmoid because logistic regression is already naturally sigmoid itself \cite{Coin}. The linear function it uses is determined through randomized weights that are transposed and multiplied by a set of x matrices containing our feature vectors. That being said, the sigmoid function is presented in equation 3.10, and its $z$ value is the equation $w^Tx$. \par
    \begin{equation}
        \sigma(z) = \frac{1}{1+e^{-z}}
    \end{equation}
    Before making a classification, a comparison is made between the value determined by the sigmoid function in equation 9 and a threshold of $0.5$, which in the case of sigmoid, is the value exactly halfway between 0 and 1, which also happen to be our targets where 0 is benign, and 1 is malware. In essence, if the value calculated by sigmoid is $> 0.5$, we classify a sample as malicious; if sigmoid is $< 0.5$, then we predict the sample to be benign. This comparison is seen in equation 3.10, and a graph of Logistic regression can be observed in figure 3.8. \par
    
    \begin{figure}[!ht]
        \begin{center}
            \hbox{\hspace{24.5ex} \includegraphics[width=.45\textwidth]{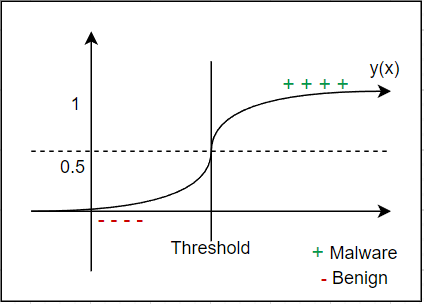}}
            \caption{\large{Logistic Regression Model}}
        \end{center}
    \end{figure}
    
    \begin{equation}
        \begin{cases}
            \sigma(z) < 0.5 \textrm{ predict 0 (benign)} \\
            \sigma(z) \geq 0.5 \textrm{ predict 1 (malware)}
        \end{cases}
    \end{equation}
    
    Using this Logistic Regression model, we managed to predict malware samples with an accuracy of 72.5\%. In addition, we achieved a precision and recall of 84\% and 72.5\%, respectively. Finally, we determined an F1-score of 77.9\% from these two values, making logistic regression our second lowest-rated model. \par
    
    \section{\large{Explainable Machine Learning}}
    
    \tab The detection of malicious software has widely advanced throughout the years. As a result, it has become more challenging to detect and for adversaries to create new types of malware that might bypass a machine learning or deep learning model especially if a malware sample is not in existing databases. These zero-day samples are introduced everyday. McAfee's advanced threat research report for 2020 was recently released in April 2021 to detail their findings against the security of computer systems. They found that from the third quarter of 2020 to the fourth quarter, new malware appeared at a rate of 26.84\% as compared to the previous quarter \cite{McAfee}. With this cat and mouse game of advancements from both sides, the complexity of algorithms begins to become too difficult for the average person to understand since all a piece of software will do is classify without explaining the results or pieces of data that lead to that classification \cite{explain}. Explainable machine learning is usually based on simple models such as linear regression, decision trees, and Naive-Bayes \cite{explain}. We cover a similar case study in the form of OneR in an attempt to examine robust classification. The algorithm creates rules based on a training set where target classes might repeat, if they do repeat for the same number in one of the feature vectors, then a rule is created in this place, and the values before this point do not matter. Without visualizing data, a typical computer user most likely will not understand what is happening. Fortunately, OneR is both simple and easy to understand for users outside of the realm of computer science. Usually, the algorithm follows a specific set of steps, as shown in figure 3.9. \par
    
    \begin{figure}[!ht]
        \begin{center}
            \hbox{\includegraphics[width=1\textwidth]{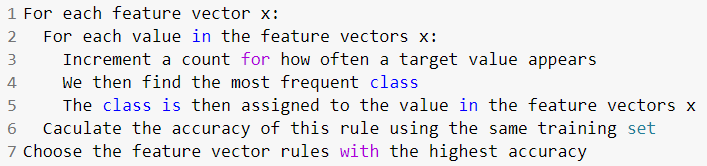}}
            \caption{\large{Original OneR}}
        \end{center}
    \end{figure}
    
    In our recreation of OneR, we create more than one rule per target attribute, and being that our feature vectors are in a numerical form, our rules will be in interval form. The basic premise is that because the data in our dataset jumps around between multiple values, we need more than one majority rule per target class. For example, if we take an interval and  set a rule value where if our feature vector falls within its range, then we classify the sample as benign, then the classification would be incorrect for a large chunk of malware samples simply because there was one more malware samples in our training set than benign samples. Although difficult to understand, this explanation could be broken down by referencing table 3.2 and classifying a new instance within the given interval: $1,102,456 < x \leq 5,944,750$. It is worth noting that this interval is only relevant for the feature vector, cyclesct. Additionally, this table is not a literal section from our training dataset, it was created to help explain how OneR limits itself by only making one rule based on the majority class. \par
    
    \begin{table}[H]
        \caption{\large{Potential Rule Intervals}}
        \begin{center}
            \hbox{\hspace{29ex} \includegraphics[width=.35\textwidth]{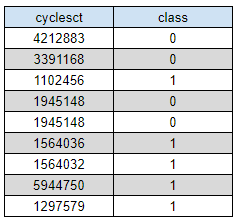}}
        \end{center}
    \end{table}
    
    If the OneR model attempts to classify a benign sample with a cyclesct value of $1,200,00$, the model would incorrectly give it a value of 1 that signifies malware. The algorithm could avoid an incorrect classification if the model created many rules while still being finite instead of only one rule. In figure 3.2, an additional rule can be created where the values repeat. In this case, they repeat at rows 4 and 5, so the interval is broken in two at this point. That is, the rules become $\textrm{if }1,102,456 < x \leq 1,945,148 \textrm{then class 0}$ and $\textrm{if }1,945,149 < x \leq 5,944,750  \textrm{ then class 1}$. We took the  many rule type of approach for the entire training dataset. By doing so, the model created between 15 and 25 rules depending on the feature vector, which includes nodeloads, dTLBstores, CPU/branch-instructions, and cyclesct. \par
    
    Our model chose node-loads as its feature vector due to it having a higher accuracy of 85.28\%. After this was determined, a base case was set at the value 0. If node-loads equals zero, then we determine the sample to be benign. Next, 24 total rules were created in interval form as mentioned above; a small sample of these rules is shown in table 3.3. They were created by artificially cutting the data at any point where a value repeated itself. The test samples were tested against these rules for classification. Finally, we created both an OneR model and a modified OneR, explained in the preceding paragraphs. By doing so, we could adequately compare the accuracy of each model and make a proper determination of which model is more efficient. \par
    
    \begin{table}[H]
        \caption{\large{First 10 Sample Rules}}
        \begin{center}
            \hbox{\hspace{29ex} \includegraphics[width=.35\textwidth]{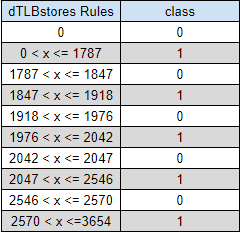}}
        \end{center}
    \end{table}
    
    The original OneR model ran the test data with an accuracy of 58.6\%. Our modified OneR increased this accuracy up to 95.99\%, a difference of 37.39\%, which is a significant difference. The model's output gives any user a proper comparison of each feature vector, explaining why node-loads are chosen. Additionally, each rule is printed along with its class value. If the user wanted to take a sample and classify it themselves from these rules, they would be able to, and doing so would help understand the classification results. Finally, a user is given four metrics which include the accuracy of our model. The model also returned a precision of 78.83\% and a recall of 91.23\%. From there, we were able to calculate an F1 score of 84.58\%.
    
    \section{\large{Hardware Implementation and Acceleration}}
    
    \tab As the above points illustrated, with every advancement made in machine learning and deep learning, so too does the complicated nature of malware evolve. Our goal here is to evaluate the hardware implementation of machine learning algorithms by converting them into a hardware description language (HDL) through a tool named Xilinx Vitis HLS version 2021.1. The work in  \cite{Vitis} uses a similar Vivado tool to accelerate the Sobel Algorithm used for edge detection in image processing techniques. Our performance evaluation was determined after a series of steps consisting of the MATLAB coding environment, a couple of applications that allowed us to reference machine learning and deep learning algorithms, and finally, the Vitis HLS tool. We wrote and attempted to convert seven algorithms which include KNN, MLP, OneR, bagged trees, decision tree, logistic regression, and SVM. \par
    
    \begin{figure}[!ht]
        \begin{center}
            \hbox{\hspace{0ex} \includegraphics[width=1\textwidth]{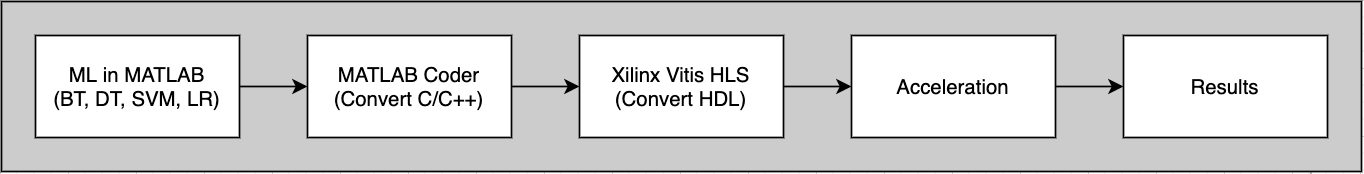}}
            \caption{\large{Hardware Implementation}}
        \end{center}
    \end{figure}
    
    Figure 3.10 provides our hardware implementation framework in which the first step was writing our different algorithms through the MATLAB software. Initially, we attempted to directly convert the seven different algorithms that we wrote in C++ into HDL by feeding them to Vitis HLS. This allowed us to encounter the issues with infinitely sized containers and unrecognized libraries that prevent the conversion of our original algorithms. Our solution led us to using MATLAB and its existing functions. This provided us support for decision trees, bagged trees, support vector machine, logistic regression, and KNN. However, the KNN source code from MATLAB provided infinite containers which meant that HDL would not be able to support this algorithm as well.
    
    The next step required the use of the MATLAB Coder application which will automatically convert our MATLAB programs in C source files. There were several options that we had control over. For example, the application allowed us to switch back and forth between infinite containers and static containers. As mentioned, Vitis HLS only accepts static containers and we therefore had that option enables in each of our experiments. The functions that MATLAB supports are also converted into C methods which meant that we did not have to worry about the creation of unsupported C libraries which would cause the HDL conversion to fail.
    
    From this point, the goal was to convert our C source files into HDL. 
    We adopted the recent implementation of HLS tool in the form of Vitis HLS 2021 version, which had a very similar GUI. The difficult learning curve for using the software was overcome by referencing the manual provided by Xilinx. The first step was creating a new project to upload our C source files and add a path for the header files. We store the project in the Xilinx directory along with any future C conversions to HDL that we create. A path for the header files needs to be inputted for each source file via the "-cFlags" option of our upload pop-up screen. Before moving on, a target board needs to be chosen; essentially, the HDL output of our program will be based on the board that we choose. In our case, we decided on a Zynq-7000 Soc FPGA board. From this point on, we can go directly into the conversion of each program. Using the "C-Synthesis" tool that Vitis HLS provides, we allow the application to create a complete transformation on its own. If there are any errors or warnings, Vitis tells us what they are through the included command line. Until these errors are fixed, we cannot continue with the conversion, and this is how we learned that vectors without a predetermined size and specific C libraries do not have an HDL equivalent.
    
    After the conversion was complete, Vitis HLS provided us with many tables that contained data such as latency values in multiple different units and a variety of resource values. Our performance and latency estimations can be gathered from these charts and we noticed that decision tree performed the best with latency and resource estimate value at 0 and 10,000 units, respectively. On the other hand, in terms of the worst resource estimates, SVM had the worst results. This was different from our initial hypothesis as we believed that bagged trees would use the most resources because the algorithm creates many decision trees and trains each of them to retrieve an average accuracy.
    
    \newpage
    \thispagestyle{plain}
    
    \chapter{\large{Results and Analysis}}
    \section{\large{Metrics}}
    
    \tab In this thesis, all of our machine learning models are evaluated against one another through accuracy, precision, recall, F-score, and finally receiver operating characteristic (ROC) area-under-curve (AUC) estimations. Accuracy ultimately denotes the rate at which a model correctly classifies a sample. These five evaluation criteria each require values from a machine learning confusion matrix to calculate their values. A confusion matrix uses four values to help understand the results of a model. The true positive (TP) metric represents the total number of correctly classified samples as true or 1 (malware). The false positive (FP) metric represents the total number of samples incorrectly classified as true or 1 (malware). The true negative (TN) metric represents the total number of samples correctly classified as false or 0 (benign). Finally, the false negative (FN) metric represents the total number of incorrectly classified samples as false or 0 (benign). If each of these metrics' values were to be summed, they would accumulate to be the total number of samples in our test set. \\
    - Accuracy: This is determined by weighing the summed number of TP and TN samples against the total samples in the test set. Generally, an algorithm's accuracy is a good measure of success unless we are given heavily imbalanced datasets. In the case of malware classification, this is usually the case. This is because there are a vast amount of benign samples when compared against malware samples.
    \begin{equation}
        Accuracy = \frac{TP + TN}{TP + FP + TN + FN}
    \end{equation}
    - Precision: Represents the probability that a classifier will predict a positive sample to be positive. It is determined by weighing the TP samples against all samples determined to be positive, even if some of those samples are incorrectly classified.
    \begin{equation}
        Precision = \frac{TP}{TP + FP}
    \end{equation}
    - Recall: Also known as Sensitivity or the true positive rate (TPR), represents the percentage of positive samples that are correctly labeled. Specifically, it is the ratio between true positives against true positives and false negatives (total of actual positive samples). This value is needed for the calculation of ROC AUC.
    \begin{equation}
        Recall (Sensitivity/TPR) = \frac{TP}{TP + FN}
    \end{equation}
    F1 Score: An F1 score measures the performance of a classifier by using the classifier's precision and recall values. It is considered a better measure for evaluation because it can take into account an imbalanced data set, such as the data sets often used in malware classification.
    \begin{equation}
        F1\:Score = 2\:\frac{Precision \times Recall}{Precision + Recall}
    \end{equation}
    ROC AUC: A ROC AUC graph allows us to measure the robustness of a classifier. It uses the TPR along the vertical axis and a false positive rate (FPR) along the horizontal axis to create the graph. The different metrics use several thresholds to get the several points that will be graphed.
    \begin{equation}
        FPR = \frac{FP}{FP + TN}
    \end{equation}
    \noindent\begin{minipage}{.5\linewidth}
        \begin{equation}
            AUC = \int_{0}^{1} TPR \, dx
        \end{equation}
    \end{minipage}%
    \begin{minipage}{.5\linewidth}
        \begin{equation}
            AUC = \int_{0}^{1} FPR \, dy
        \end{equation}
    \end{minipage}%
    
    \section{\large{Model Evaluation}}
    
    \tab In terms of accuracy, our machine learning, and deep learning models had successful results when it came to zero-day malware detection accuracy. However, we were running imbalanced data sets through the models, which is typical for malware classification. Often is the case where new types of malware arise, and there is not yet enough information to create a balanced data set to go against the more common benign samples. Therefore, it was more feasible to compare F1 scores, which consider both the precision and recall values and not the returned accuracies of the models, which do not consider the imbalance. Nonetheless, Table 4 contains the values of each metric for all seven models that we created.
    
    Zero-day data has proved to have less successful results in standard ML-based algorithms compared to ensemble learning techniques \cite{ISQED}. Most of these ML-based techniques are incapable of producing models in the 90th percentile of accuracy. Two of our seven algorithms managed to score above 90\%, another three scored above 80\%, and the final two scored under a 78\%, all while running zero-day data. Our best model was a decision tree that, based on the data used, managed to classify 91.2\% of samples correctly and produced an F1-score of 91.5\%. Notably, the only changes made to our zero-day data were the same feature engineering based on mutual information, which was also conducted for our training data. An accuracy and F1-score as high as the two achieved proves to be around average for classifiers predicting the class of zero-day data. If these values were above 95\% and even closer to 99\%, then the zero-day models would have proven successful; however, an accuracy or F1 score that large requires more advanced algorithms rather than standard ML.
    
    \begin{table}
        \caption{\large{Model Metrics}}
        \begin{center}
            \hbox{\hspace{0ex} \includegraphics[width=1\textwidth]{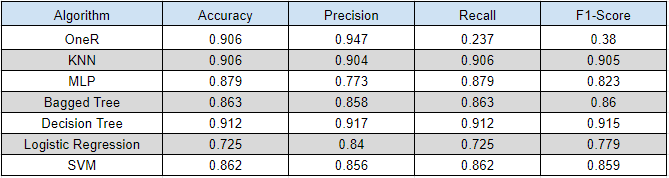}}
        \end{center}
    \end{table}
    
    We also used ROC AUC graphs as a secondary metric of an algorithm's robustness or ability to classify binary benign and malware samples. An algorithm with an AUC above a .5 can be considered a successful implementation of an algorithm, while one with a .5 or lower might be doing nothing more but guessing the binary class of a sample. An AUC of .5 means that an algorithm had 50\% true positive and 50\% false-negative, while an AUC of 1 (best) means that the model had 100\% true positives. In figure 9 and table 5, we combined the graphs/values of each algorithm for a better comparison of each. We can see that the decision tree was not only the most accurate and best algorithm for imbalanced data sets, but it was also the most robust as it produced an AUC of 1.0. This was followed by OneR and then KNN and directly correlates to the order of accuracy from table 4.1.
    
    \begin{figure}[!ht]
        \begin{center}
            \hbox{\hspace{13ex} \includegraphics[width=.65\textwidth]{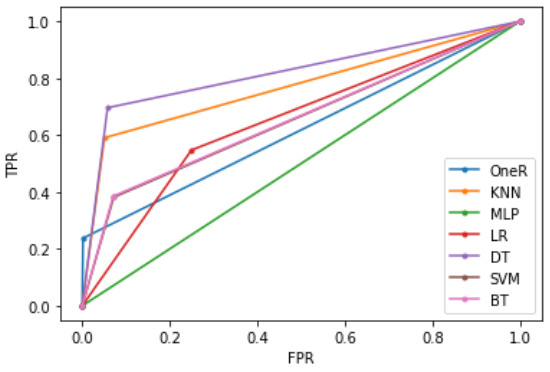}}
            \caption{\large{ROC AUC Graphs}}
        \end{center}
    \end{figure}
    
    \begin{table}[H]
        \caption{\large{AUC Values}}
        \begin{center}
            \hbox{\hspace{22ex} \includegraphics[width=.5\textwidth]{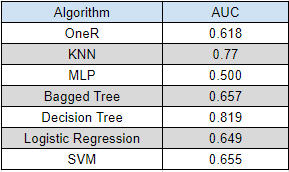}}
        \end{center}
    \end{table}
    
    \section{\large{Hardware Evaluation}}
    
    \tab After converting C source files to HDL, Vitis HLS gives us a few different tables containing performance parameters from our algorithms. One such value was for latency which measures the delay that our acceleration experienced. The synthesis provided latency in terms of total cycles and total time in nanoseconds. We hypothesized that bagged trees would have the most considerable overhead from all of our models due to having to run multiple loops that would create the numerous decision trees used to make classifications. This was not the case, figure 4.2 and 4.3 show that SVM experienced the most considerable overhead with a latency of 52 cycles and 1300 nanoseconds while bagged trees experienced the second-lowest. Notably, the decision tree did not experience any notable latency compared to every other algorithm and therefore had a bar near zero in our charts. 
    
    \begin{figure}[!ht]
        \begin{center}
            \hbox{\hspace{18ex} \includegraphics[width=.6\textwidth]{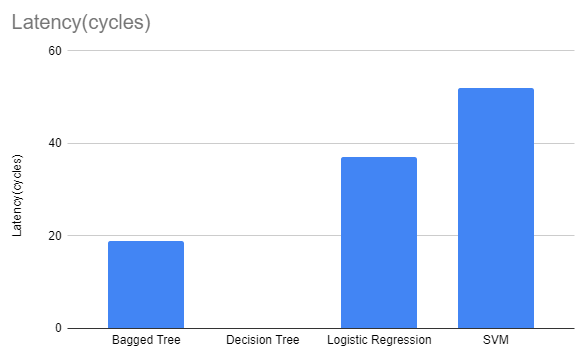}}
            \caption{\large{Latency (Cycles)}}
        \end{center}
    \end{figure}
    
    \begin{figure}[!ht]
        \begin{center}
            \hbox{\hspace{18ex} \includegraphics[width=.6\textwidth]{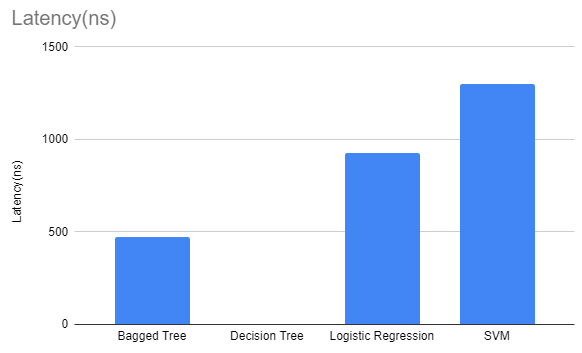}}
            \caption{\large{Latency (Nanoseconds)}}
        \end{center}
    \end{figure}
    
    In this case, SVM reached its most significant latency on interval 53. Interval refers to the total number of clock cycles before new inputs can be applied. It occurs after the final latency cycle (52), and our acceleration continues once the final output is written. Similarly, the latency in nanoseconds is the total time before the acceleration could continue. It has an identical ratio to latency in cycles meaning that when the final process completes, that is also when the total time has been met before the next interval continues. \par
    
    Furthermore, the synthesis also returned resource estimates for each algorithm. These estimates were for BRAM, DSP, FF, and LUT. When summed together, we receive the total Model Resource Estimate (MRE) that accurately represents the overhead that our models experienced. According to Xilinx, BRAM (Block Ram) specifies the block RAM utilization, DSP specifies the DSP utilization, FF refers to the utilized registers, and finally, LUT (LookUp Table) refers to the utilization of the LUT. In figure 4.4, we have summed the total RME for each model and see that Bagged Trees experienced the most significant overhead.
    
    \begin{figure}[H]
        \begin{center}
            \hbox{\hspace{18ex} \includegraphics[width=.6\textwidth]{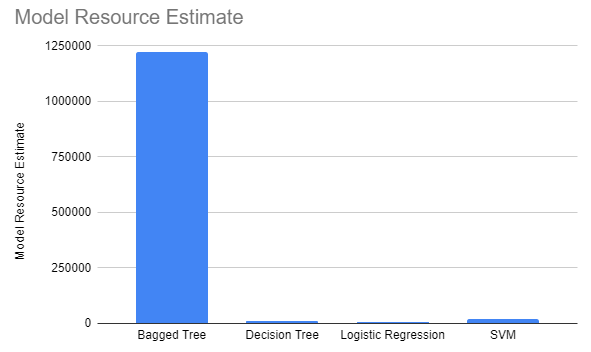}}
            \caption{\large{Resource Model Estimate (RME)}}
        \end{center}
    \end{figure}
    
    \newpage
    \thispagestyle{plain}

    \chapter{\large{Conclusion and Future Work}}
    
    \tab\\ \tab\tab In this thesis, we covered multiple topics which included explainable machine learning, efficiency analysis and evaluating various machine learning algorithms for hardware-based cybersecurity, and finally analyzing the performance overhead and resource estimates of machine learning algorithms when deployed on the hardware. The optimization of machine learning and deep learning algorithms can be the next vital step at detecting malware. With algorithms having relatively low overhead and latency cycles/times, a successful approach can be conducted to encompass most if not all techniques. Moreover, with machine learning moving towards an explainable approach, users gain the ability to understand and interpret the fundamentals and details of an algorithm and are not left in the dark about their classification results. This becomes even a more important point of improvement when it comes to application of machine learning for security-critical applications. Combined with the acceleration of the algorithms, this can lead to a user preventing a malicious file from even getting onto their computers or hardware project boards. We have thoroughly compared the efficiency of various types of machine learning algorithms for hardware-based malware detection across different  metrics such accuracy, F-score, AUC, area overhead, and latency. Our experimental results highlight the effectiveness of decision tree algorithm in terms of detection accuracy. Also, by taking the impact of hardware implementation cost into consideration algorithm decision tree also delivered the most efficient results. Based on our metrics used, decision tree performed with an accuracy of 91.2\%, and an F-score of 91.5\%. In terms of area under the curve, the algorithm proved to have a robustness equal to 81.9\%. Finally, when it comes to overhead, decision tree did not experience any substantial latency but did use a minimal amount of resources at 10,328 units. 
    
    As our future direction, we would like to expand on the focal points we covered by examining more number of machine learning algorithms for developing efficient hardware-based cybersecurity countermeasures. Regarding explainable machine learning, decision trees and bayesian network models are the algorithms that could be used due to their their simplicity when visualizing how classifications occur. However, modern malware detection techniques use neural networks which leads to an overall increased accuracy from classifications. The difficulty is that neural nets are difficult to explain and visualize; therefore, we would like to implement a Naive-Bayes and decision tree classifier to escape the black box caused by deep learning. A Naive-Bayes classifier uses each feature independently and creates probability values of each column while ignoring any likelihood of one feature affecting one of the others. Fortunately, explaining how the algorithm computes the probability values is much more simple in comparison to explaining the net that connects nodes together in a deep learning model. Additionally, we would like to expand our hardware implementation framework to a wider number of algorithms across different machine learning branches. 
    
    As another direction, we would like to explore the suitability and hardware implementations of machine learning algorithms for detecting micorarchitectural Side-Channel Attacks (SCAs). Microarchitectural SCAs have also posed serious threats to the security of modern computing systems. Such attacks exploit side-channel vulnerabilities originating from fundamental performance improving components such as cache memories \cite{SCA}. Therefore, examining the cost-efficiency of different standard and advanced machine learning and deep learning algorithms for SCAs detection could be another important direction to expand this research.
    
    \newpage
    \thispagestyle{plain}
    \renewcommand\refname{\begin{center} \fontsize{12pt}{12pt}\uppercase{\Large\textbf{References}} \end{center}}
    \printbibliography

@inproceedings{cache,
        author = {Zhang, Tianwei and Zhang, Yinqian and Lee, Ruby B.},
        title = {Analyzing Cache Side Channels Using Deep Neural Networks},
        year = {2018},
        isbn = {9781450365697},
        publisher = {Association for Computing Machinery},
        address = {New York, NY, USA},
        url = {https://doi.org/10.1145/3274694.3274715},
        doi = {10.1145/3274694.3274715},
        booktitle = {Proceedings of the 34th Annual Computer Security Applications Conference},
        pages = {174–186},
        numpages = {13},
        location = {San Juan, PR, USA},
        series = {ACSAC '18}
    }

@ARTICLE{ML_DL,
  author={Xin, Yang and Kong, Lingshuang and Liu, Zhi and Chen, Yuling and Li, Yanmiao and Zhu, Hongliang and Gao, Mingcheng and Hou, Haixia and Wang, Chunhua},
  journal={IEEE Access}, 
  title={Machine Learning and Deep Learning Methods for Cybersecurity}, 
  year={2018},
  volume={6},
  number={},
  pages={35365-35381},
  doi={10.1109/ACCESS.2018.2836950}}

@ARTICLE{Robust,
  author={Vinayakumar, R. and Alazab, Mamoun and Soman, K. P. and Poornachandran, Prabaharan and Venkatraman, Sitalakshmi},
  journal={IEEE Access}, 
  title={Robust Intelligent Malware Detection Using Deep Learning}, 
  year={2019},
  volume={7},
  number={},
  pages={46717-46738},
  doi={10.1109/ACCESS.2019.2906934}}

@article{Security,
  author    = {Yoni Birman and
               Shaked Hindi and
               Gilad Katz and
               Asaf Shabtai},
  title     = {{ASPIRE:} Automated Security Policy Implementation Using Reinforcement
               Learning},
  journal   = {CoRR},
  volume    = {abs/1905.10517},
  year      = {2019},
  url       = {http://arxiv.org/abs/1905.10517},
  eprinttype = {arXiv},
  eprint    = {1905.10517},
  bibsource = {dblp computer science bibliography, https://dblp.org}
}

@article{Detection,
        author = {Yuxin, Ding and Siyi, Zhu},
        title = {Malware detection based on deep learning algorithm},
        year = {2019},
        publisher = {Neural Computing and Applications},
        url = {https://doi.org/10.1007/s00521-017-3077-6},
        doi = {10.1007/s00521-017-3077-6},
        pages = {461-472},
        numpages = {11}
    }

@misc{
Lazy_Eager,
title={An Exhaustive Analysis of Lazy vs. Eager Learning Methods for Real-Estate Property Investment},
author={Setareh Rafatirad and Maryam Heidari},
year={2019},
url={https://openreview.net/forum?id=r1ge8sCqFX},
}

@inproceedings{micro,
author = {Dinakarrao, Sai Manoj Pudukotai and Amberkar, Sairaj and Bhat, Sahil and Dhavlle, Abhijitt and Sayadi, Hossein and Sasan, Avesta and Homayoun, Houman and Rafatirad, Setareh},
title = {Adversarial Attack on Microarchitectural Events Based Malware Detectors},
year = {2019},
isbn = {9781450367257},
publisher = {Association for Computing Machinery},
address = {New York, NY, USA},
url = {https://doi.org/10.1145/3316781.3317762},
doi = {10.1145/3316781.3317762},
booktitle = {Proceedings of the 56th Annual Design Automation Conference 2019},
articleno = {164},
numpages = {6},
location = {Las Vegas, NV, USA},
series = {DAC '19}
}

@ARTICLE{Coin,
  author={Liu, Wenye and Chang, Chip-Hong and Wang, Xueyang and Liu, Chen and Fung, Jason M. and Ebrahimabadi, Mohammad and Karimi, Naghmeh and Meng, Xingyu and Basu, Kanad},
  journal={IEEE Journal on Emerging and Selected Topics in Circuits and Systems}, 
  title={Two Sides of the Same Coin: Boons and Banes of Machine Learning in Hardware Security}, 
  year={2021},
  volume={11},
  number={2},
  pages={228-251},
  doi={10.1109/JETCAS.2021.3084400}}

@article{HPC1,
author = {Kuruvila, Abraham Peedikayil and Mahapatra, Anushree and Karri, Ramesh and Basu, Kanad},
title = {Hardware Performance Counters: Ready-Made vs Tailor-Made},
year = {2021},
issue_date = {October 2021},
publisher = {Association for Computing Machinery},
address = {New York, NY, USA},
volume = {20},
number = {5s},
issn = {1539-9087},
url = {https://doi.org/10.1145/3476996},
doi = {10.1145/3476996},
journal = {ACM Trans. Embed. Comput. Syst.},
month = sep,
articleno = {65},
numpages = {26}
}

@ARTICLE{Defense,
  author={Kuruvila, Abraham Peedikayil and Kundu, Shamik and Basu, Kanad},
  journal={IEEE Transactions on Computer-Aided Design of Integrated Circuits and Systems}, 
  title={Defending Hardware-Based Malware Detectors Against Adversarial Attacks}, 
  year={2021},
  volume={40},
  number={9},
  pages={1727-1739},
  doi={10.1109/TCAD.2020.3026960}}

@INPROCEEDINGS{explain,
  author={Pan, Zhixin and Sheldon, Jennifer and Mishra, Prabhat},
  booktitle={2020 IEEE 38th International Conference on Computer Design (ICCD)}, 
  title={Hardware-Assisted Malware Detection using Explainable Machine Learning}, 
  year={2020},
  volume={},
  number={},
  pages={663-666},
  doi={10.1109/ICCD50377.2020.00113}}

@INPROCEEDINGS{PMC,
  author={Pattee, Jordan and Lee, Byeong Kil},
  booktitle={2020 IEEE 39th International Performance Computing and Communications Conference (IPCCC)}, 
  title={Design Alternatives for Performance Monitoring Counter based Malware Detection}, 
  year={2020},
  volume={},
  number={},
  pages={1-2},
  doi={10.1109/IPCCC50635.2020.9391559}}

@article {Kwan,
  author={Kwan, Abigail},
  title = {Malware Detection at the Microarchitecture Level Using Machine Learning Techniques},
  year = {2020},
  numpages = {35}}

@inproceedings{rao1997visualizing,
  title={Visualizing Bagged Decision Trees.},
  author={Rao, J Sunil and Potts, William JE},
  booktitle={KDD},
  pages={243--246},
  year={1997}
}

@INPROCEEDINGS{Vitis,
  author={Mousouliotis, Panagiotis and Zogas, Stavros and Christakos, Panagiotis and Keramidas, Georzios and Petrellis, Nikos and Antonopoulos, Christos and Voros, Nikolaos},
  booktitle={2021 10th Mediterranean Conference on Embedded Computing (MECO)}, 
  title={Exploiting Vitis Framework for Accelerating Sobel Algorithm}, 
  year={2021},
  volume={},
  number={},
  pages={1-5},
  doi={10.1109/MECO52532.2021.9460221}}

@article{Botnets,
  title={Turning internet of things (iot) into internet of vulnerabilities (iov): Iot botnets},
  author={Angrishi, Kishore},
  journal={arXiv preprint arXiv:1702.03681},
  year={2017}
}

@article{spyware,
  title={Dynamic spyware analysis},
  author={Egele, Manuel and Kruegel, Christopher and Kirda, Engin and Yin, Heng and Song, Dawn},
  year={2007},
  publisher={Advanced Computing Systems Professional and Technical Association}
}

@inproceedings{sayadi_2smart:_2019,
	address = {Florence, Italy},
	title = {{2SMaRT}: {A} {Two}-{Stage} {Machine} {Learning}-{Based} {Approach} for {Run}-{Time} {Specialized} {Hardware}-{Assisted} {Malware} {Detection}},
	isbn = {9783981926323},
	shorttitle = {{2SMaRT}},
	url = {https://ieeexplore.ieee.org/document/8715080/},
	doi = {10.23919/DATE.2019.8715080},
	urldate = {2020-04-18},
	booktitle = {2019 {Design}, {Automation} \& {Test} in {Europe} {Conference} \& {Exhibition} ({DATE})},
	publisher = {IEEE},
	author = {Sayadi, Hossein and Makrani, Hosein Mohammadi and Pudukotai Dinakarrao, Sai Manoj and Mohsenin, Tinoosh and Sasan, Avesta and Rafatirad, Setareh and Homayoun, Houman},
	month = mar,
	year = {2019},
	pages = {728--733}
}

@inproceedings{sayadi_ensemble_2018,
	address = {San Francisco, CA},
	title = {Ensemble {Learning} for {Effective} {Run}-{Time} {Hardware}-{Based} {Malware} {Detection}: {A} {Comprehensive} {Analysis} and {Classification}},
	isbn = {9781538641149},
	shorttitle = {Ensemble {Learning} for {Effective} {Run}-{Time} {Hardware}-{Based} {Malware} {Detection}},
	url = {https://ieeexplore.ieee.org/document/8465828/},
	doi = {10.1109/DAC.2018.8465828},
	urldate = {2020-04-27},
	booktitle = {2018 55th {ACM}/{ESDA}/{IEEE} {Design} {Automation} {Conference} ({DAC})},
	publisher = {IEEE},
	author = {Sayadi, Hossein and Patel, Nisarg and P.D., Sai Manoj and Sasan, Avesta and Rafatirad, Setareh and Homayoun, Houman},
	month = jun,
	year = {2018},
	pages = {1--6}
}

@inproceedings{sayadi2018customized,
  title={Customized machine learning-based hardware-assisted malware detection in embedded devices},
  author={Sayadi, Hossein and Makrani, Hosein Mohammadi and Randive, Onkar and PD, Sai Manoj and Rafatirad, Setareh and Homayoun, Houman},
  booktitle={2018 17th IEEE International Conference On Trust, Security And Privacy In Computing And Communications/12th IEEE International Conference On Big Data Science And Engineering (TrustCom/BigDataSE)},
  pages={1685--1688},
  year={2018},
  organization={IEEE}
}

@inproceedings{Sayadi-glsvlsi20,
author = {Sayadi, Hossein and et al.},
title = {StealthMiner: Specialized Time Series Machine Learning for Run-Time Stealthy Malware Detection Based on Microarchitectural Features},
year = {2020},
booktitle = {GLSVLSI'20},
pages = {175–180},
numpages = {6},
location = {Virtual Event, China},
}

@inproceedings{Demme-ISCA13,
 author = {Demme, J. and et al.},
 title = {On the Feasibility of Online Malware Detection with Performance Counters},
 booktitle = {ISCA'13},
 year = {2013},
 pages = {559--570},
 publisher = {ACM},
}

@InProceedings{Tang,
author="Tang, A.
and et al.",
title="Unsupervised Anomaly-Based Malware Detection Using Hardware Features",
  author={Tang, Adrian and Sethumadhavan, Simha and Stolfo, Salvatore J},
  booktitle={RAID'14},
  pages={109--129},
  year={2014},
  organization={Springer}
}

@inproceedings{Rootkit-Singh,
 author = {Singh, B. and et al.},
 title = {On the Detection of Kernel-Level Rootkits Using Hardware Performance Counters},
 booktitle = {ASIACCS'17},
 year = {2017},
 location = {Abu Dhabi, United Arab Emirates},
 pages = {483--493},
}

@article{sayadi2021towards,
  title={Towards Accurate Run-Time Hardware-Assisted Stealthy Malware Detection: A Lightweight, Yet Effective Time Series CNN-Based Approach},
  author={Sayadi, Hossein and Gao, Yifeng and Makrani, Hosein Mohammadi and Lin, Jessica and Costa, Paulo Cesar and Rafatirad, Setareh and Homayoun, Houman},
  journal={Cryptography},
  volume={5},
  number={4},
  pages={28},
  year={2021},
  publisher={Multidisciplinary Digital Publishing Institute}
}

@INPROCEEDINGS{adaptive-HMD,
  author={Gao, Yifeng and Makrani, Hosein Mohammadi and Aliasgari, Mehrdad and Rezaei, Amin and Lin, Jessica and Homayoun, Houman and Sayadi, Hossein},
  booktitle={2021 IEEE 27th International Symposium on On-Line Testing and Robust System Design (IOLTS)}, 
  title={Adaptive-HMD: Accurate and Cost-Efficient Machine Learning-Driven Malware Detection using Microarchitectural Events}, 
  year={2021},
  volume={},
  number={},
  pages={1-7},
  doi={10.1109/IOLTS52814.2021.9486701}}

@INPROCEEDINGS{ISQED,
  author={He, Zhangying and Miari, Tahereh and Makrani, Hosein Mohammadi and Aliasgari, Mehrdad and Homayoun, Houman and Sayadi, Hossein},
  booktitle={2021 22nd International Symposium on Quality Electronic Design (ISQED)}, 
  title={When Machine Learning Meets Hardware Cybersecurity: Delving into Accurate Zero-Day Malware Detection}, 
  year={2021},
  volume={},
  number={},
  pages={85-90},
  doi={10.1109/ISQED51717.2021.9424330}}

@misc{McAfee,
  url={https://www.mcafee.com/enterprise/en-us/lp/threats-reports/apr-2021}, 
  journal={Mcafee.com}, 
  year={2021} }

@INPROCEEDINGS{SCA,
  author={Wang, Han and Sayadi, Hossein and Rafatirad, Setareh and Sasan, Avesta and Homayoun, Houman},
  booktitle={2020 IEEE 26th International Symposium on On-Line Testing and Robust System Design (IOLTS)}, 
  title={SCARF: Detecting Side-Channel Attacks at Real-time using Low-level Hardware Features}, 
  year={2020},
  volume={},
  number={},
  pages={1-6},
  doi={10.1109/IOLTS50870.2020.9159708}}

@inproceedings{mwscas_2020recent,
  title={Recent advancements in microarchitectural security: Review of machine learning countermeasures},
  author={Sayadi, Hossein and Wang, Han and Miari, Tahereh and Makrani, Hosein Mohammadi and Aliasgari, Mehrdad and Rafatirad, Setareh and Homayoun, Houman},
  booktitle={2020 IEEE 63rd International Midwest Symposium on Circuits and Systems (MWSCAS)},
  pages={949--952},
  year={2020},
  organization={IEEE}
}

@inproceedings{wang2020comprehensive,
  title={Comprehensive evaluation of machine learning countermeasures for detecting microarchitectural side-channel attacks},
  author={Wang, Han and Sayadi, Hossein and Sasan, Avesta and Rafatirad, Setareh and Mohsenin, Tinoosh and Homayoun, Houman},
  booktitle={Proceedings of the 2020 on Great Lakes Symposium on VLSI},
  pages={181--186},
  year={2020}
}

@inproceedings{sayadi2017-igsc,
  title={Scheduling multithreaded applications onto heterogeneous composite cores architecture},
  author={Sayadi, Hossein and Homayoun, Houman},
  booktitle={2017 Eighth International Green and Sustainable Computing Conference (IGSC)},
  pages={1--8},
  year={2017},
  organization={IEEE}
}

@inproceedings{makrani2018energy,
  title={Energy-aware and machine learning-based resource provisioning of in-memory analytics on cloud},
  author={Makrani, Hosein Mohammadi and Sayadi, Hossein and Motwani, Devang and Wang, Han and Rafatirad, Setareh and Homayoun, Houman},
  booktitle={Proceedings of the ACM Symposium on Cloud Computing},
  pages={517--517},
  year={2018}
}

@inproceedings{wang2020mitigating,
  title={Mitigating cache-based side-channel attacks through randomization: A comprehensive system and architecture level analysis},
  author={Wang, Han and Sayadi, Hossein and Mohsenin, Tinoosh and Zhao, Liang and Sasan, Avesta and Rafatirad, Setareh and Homayoun, Houman},
  booktitle={2020 Design, Automation \& Test in Europe Conference \& Exhibition (DATE)},
  pages={1414--1419},
  year={2020},
  organization={IEEE}
}

@inproceedings{kocher2019spectre,
  title={Spectre attacks: Exploiting speculative execution},
  author={Kocher, Paul and Horn, Jann and Fogh, Anders and Genkin, Daniel and Gruss, Daniel and Haas, Werner and Hamburg, Mike and Lipp, Moritz and Mangard, Stefan and Prescher, Thomas and others},
  booktitle={2019 IEEE Symposium on Security and Privacy (SP)},
  pages={1--19},
  year={2019},
  organization={IEEE}
}

@article{lipp2018meltdown,
  title={Meltdown},
  author={Lipp, Moritz and Schwarz, Michael and Gruss, Daniel and Prescher, Thomas and Haas, Werner and Mangard, Stefan and Kocher, Paul and Genkin, Daniel and Yarom, Yuval and Hamburg, Mike},
  journal={arXiv preprint arXiv:1801.01207},
  year={2018}
}

@incollection{KIRA,
title = {A Practical Approach to Feature Selection},
editor = {Derek Sleeman and Peter Edwards},
booktitle = {Machine Learning Proceedings 1992},
publisher = {Morgan Kaufmann},
address = {San Francisco (CA)},
pages = {249-256},
year = {1992},
isbn = {978-1-55860-247-2},
doi = {https://doi.org/10.1016/B978-1-55860-247-2.50037-1},
url = {https://www.sciencedirect.com/science/article/pii/B9781558602472500371},
author = {Kenji Kira and Larry A. Rendell}
}

@article{JANGJACCARD,
title = {A survey of emerging threats in cybersecurity},
journal = {Journal of Computer and System Sciences},
volume = {80},
number = {5},
pages = {973-993},
year = {2014},
note = {Special Issue on Dependable and Secure Computing},
issn = {0022-0000},
doi = {https://doi.org/10.1016/j.jcss.2014.02.005},
url = {https://www.sciencedirect.com/science/article/pii/S0022000014000178},
author = {Julian Jang-Jaccard and Surya Nepal}
}

@INPROCEEDINGS{instance,
  author={Solomatine, D.P. and Maskey, M. and Shrestha, D.L.},
  booktitle={The 2006 IEEE International Joint Conference on Neural Network Proceedings}, 
  title={Eager and Lazy Learning Methods in the Context of Hydrologic Forecasting}, 
  year={2006},
  volume={},
  number={},
  pages={4847-4853},
  doi={10.1109/IJCNN.2006.247163}}

@article{IoT,
author = {Darabian, Hamid and Dehghantanha, Ali and Hashemi, Sattar and Homayoun, Sajad and Choo, Kim-Kwang Raymond},
title = {An opcode-based technique for polymorphic Internet of Things malware detection},
journal = {Concurrency and Computation: Practice and Experience},
volume = {32},
number = {6},
pages = {e5173},
doi = {https://doi.org/10.1002/cpe.5173},
url = {https://onlinelibrary.wiley.com/doi/abs/10.1002/cpe.5173},
eprint = {https://onlinelibrary.wiley.com/doi/pdf/10.1002/cpe.5173},
note = {e5173 cpe.5173},
year = {2020}
}

@INPROCEEDINGS{Interpred,
  author={Gilpin, Leilani H. and Bau, David and Yuan, Ben Z. and Bajwa, Ayesha and Specter, Michael and Kagal, Lalana},
  booktitle={2018 IEEE 5th International Conference on Data Science and Advanced Analytics (DSAA)}, 
  title={Explaining Explanations: An Overview of Interpretability of Machine Learning}, 
  year={2018},
  volume={},
  number={},
  pages={80-89},
  doi={10.1109/DSAA.2018.00018}}

@article{DEPREN,
title = {An intelligent intrusion detection system (IDS) for anomaly and misuse detection in computer networks},
journal = {Expert Systems with Applications},
volume = {29},
number = {4},
pages = {713-722},
year = {2005},
issn = {0957-4174},
doi = {https://doi.org/10.1016/j.eswa.2005.05.002},
url = {https://www.sciencedirect.com/science/article/pii/S0957417405000989},
author = {Ozgur Depren and Murat Topallar and Emin Anarim and M. Kemal Ciliz},
}

@INPROCEEDINGS{rajatTrojan,
  author={Chakraborty, Rajat Subhra and Narasimhan, Seetharam and Bhunia, Swarup},
  booktitle={2009 IEEE International High Level Design Validation and Test Workshop}, 
  title={Hardware Trojan: Threats and emerging solutions}, 
  year={2009},
  volume={},
  number={},
  pages={166-171},
  doi={10.1109/HLDVT.2009.5340158}}
\end{document}